\documentclass[aps,prl,reprint,nofootinbib,twocolumn,superscriptaddress,showpacs,showkeys,longbibliography]{revtex4-1}
\usepackage{eurosym}
\usepackage{amsmath,amssymb,amstext}
\usepackage{ulem} 
\usepackage[usenames,dvipsnames]{color}
\usepackage{graphicx}
\usepackage{braket}
\usepackage{natbib}
\usepackage{comment}
\usepackage{dcolumn}
\usepackage{wasysym}

\usepackage[colorlinks,bookmarks=false,citecolor=blue,linkcolor=red,urlcolor=blue]{hyperref}

\begin{document}
\title{Dissipative Magnetic Polariton Soliton}
\author{Chunyu Jia}
\affiliation{Department of Physics, Zhejiang Normal University, Jinhua, 321004, China}
\author{Rukuan Wu}
\affiliation{Department of Physics, Zhejiang Normal University, Jinhua, 321004, China}
\author{Ying Hu}
\email{The corresponding author: huying@sxu.edu.cn}
\affiliation{State Key Laboratory of Quantum Optics and Quantum Optics Devices, Institute of Laser Spectroscopy, Shanxi University, Taiyuan, Shanxi 030006, China}
\affiliation{Collaborative Innovation Center of Extreme Optics, Shanxi University, Taiyuan, Shanxi 030006, China}
\author{Wu-Ming Liu}
\affiliation{Beijing National Laboratory for Condensed Matter Physics, Institute of Physics, Chinese Academy of Sciences, Beijing 100190, China}
\affiliation{School of Physical Sciences, University of Chinese Academy of Sciences, Beijing 100190, China}
\affiliation{Songshan Lake Materials Laboratory, Dongguan, Guangdong 523808, China}
\author{Zhaoxin Liang}
\email{The corresponding author: zhxliang@gmail.com}
\affiliation{Department of Physics, Zhejiang Normal University, Jinhua, 321004, China}

\date{\today}

\begin{abstract}
Dissipative solitons are non-decaying out-of-equilibrium entities that result from double balances between gain and loss, as well as nonlinearity and dispersion. Here we describe a scenario where establishment of double balances relies on existence of multiple collective excitation channels in open-dissipative quantum systems. It differs from conventional single-channel double-balance scenario governing dissipative solitons such as dissipative Kerr solitons, in that the soliton itself arises in a decoupled excitation channel and hence coherent nonlinear dynamics, but its background state corresponds to other channels and determined by the balance of gain and loss. We demonstrate through a spinor polariton Bose-Einstein condensate (BEC) under spatially uniform non-resonant pumping, where we show the existence of dissipative magnetic matter-light soliton that represents an exact solution to two-component driven-dissipative Gross-Pitaevskii equations. It manifests as a localized spin polarization with the background state being linearly polarized, and does not decay when propagating in the dissipative medium. Our findings offer new benchmarks as well as a new route for understanding and realizing dissipative solitons.

\end{abstract}

\maketitle

Solitons are localized wave packets capable of maintaining their shape in propagation. In Hamiltonian systems, solitons originate from a single balance between dispersion and nonlinearity~\cite{HamSolitonRev1,SolitonRev1,SolitonRev0,Liang2005}. In distinct contrast, dissipative nonlinear systems typically involve gain and loss of matter or energy~\cite{Dissipation1,Dissipation2,Rev2}, where solitons usually exist within a finite lifetime before eventually vanishing. Remarkably, the composite balances of gain and dissipation, along with dispersion and nonlinearity, provides core mechanisms for forming solitons that do not decay, and so define dissipative solitons~\cite{DissipativeSoliton1,DissipativeSoliton2,DissipativeBook1,DissipativeBook2,DissipativeSolitonRev1}. Dissipative solitons are not only of fundamental interests as a nonlinear dynamics problem \textit{per se} or as intriguing collective excitations in far-from-equilibrium quantum matter, but also provide potential resources for practical applications~\cite{DissipativeSolitonRev1,InformationProcess,InformationProcessRev1,InformationProcess2017} such as in optics and information processing where dissipative effect cannot be ignored. In the quest of dissipative solitons, exact solutions play important roles as they provide some benchmarks for understanding generic physical mechanisms behind soliton formation in various dissipative systems. However, exact solutions are few and far between, and it is challenging to search for exact dissipative soliton solutions that shed light on new scenarios to establish double balances in dissipative nonlinear systems - the latter, reciprocally, opens the route toward more new dissipative solitons. 

\begin{figure}
  \includegraphics[width=0.85\columnwidth]{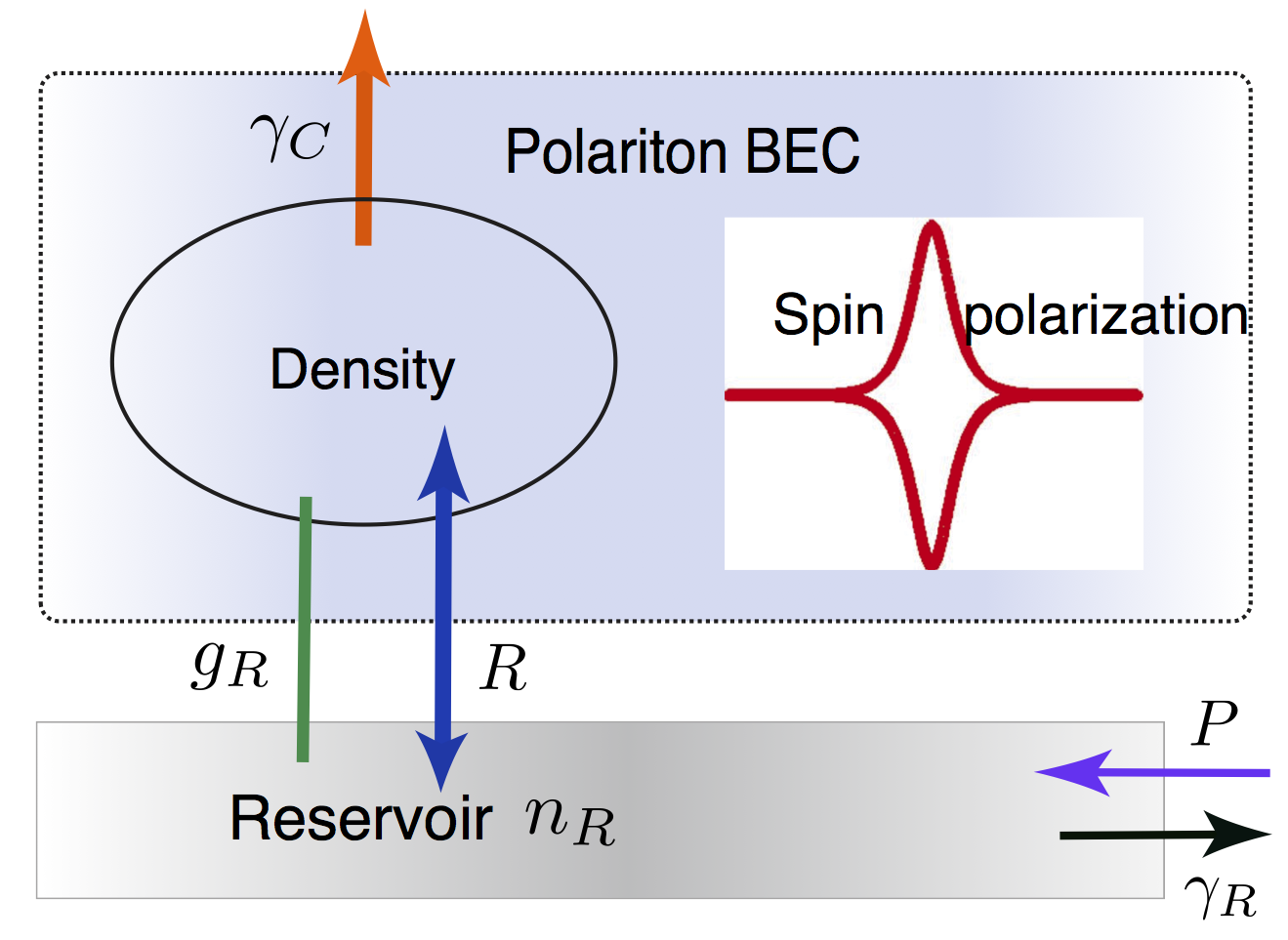}\\
\caption{Schematics of bi-channel double-balance scenario for producing dissipative magnetic soliton in a spinor polariton BEC under nonresonant pumping. Condensed polaritons with loss rate $\gamma_C$ are continuously replenished from reservoir polaritons at rate $R$. Reservoir polaritons decay at rate $\gamma_R$ while subjected to a uniform pumping $P$. Polaritons in the reservoir and condensate also interact with each other modeled by $g_R$. On top of the steady state, there exist two excitation channels - density and spin polarization - which are usually coupled. When the gain balances loss fixing the background density, nonlinear spin-polarization dynamics takes place in an isolated channel rendering a soliton profile that manifests a balance of nonlinearity and dispersion. 
}\label{figure1}
\end{figure}

In this Letter, we describe and analyze a novel type of vector dissipative soliton - referred to as dissipative magnetic soliton, which represents an exact solution to driven-dissipative two-component Gross-Pitaevskii (GP) equation that describes a spinor polariton Bose-Einstein condensates (BEC) formed under spatially uniform nonresonant pumping~\cite{Rev0,Rev1,Rev2,Rev3}. It is a localized spin polarization on top of a spin-balanced density background, which, remarkably, can move without decaying, as opposed to most matter-light solitons in polariton BECs that exist in finite lifetimes~\cite{Xue2014,Smirnov2014,Pinsker2014,Pinsker2015,Pinsker2016,Ma2017,Ma2018,Xu2019,Amo2011,Cilibrizzi2014,Walker2017}. This  exact solution allows us to reveal the mechanism underlying dissipative magnetic soliton, i.e., a bi-channel double balance [see Fig.~\ref{figure1}]: The soliton is contained in the nonlinear spin-polarization dynamics which is decoupled from the dynamics of reservoir and density of condensed polaritons and consequently coherent, with a profile manifesting a balance of nonlinearity and dispersion; on the other hand, the background state for the soliton corresponds to the spin-balanced density of condensed polaritons, which is dictated by the balance of gain and loss. This bi-channel double-balance principle makes the dissipative magnetic soliton significantly distinctive from conventional dissipative solitons in complex Ginzburg-Landau equation (see e.g. Refs.~\cite{Dissipation1,DissipativeSolitonRev1,AkhmedievPRL1995,AkhmedievPRE1996}), where composite balances occur in a single excitation channel, as well as from recently found dissipative polariton solitons, which rely on particular forms of pumping such as Gaussian(-like) nonresonant pumping~\cite{Pinsker2014,Ma2017,Ma2018} or resonant pumping~\cite{ResonantSoliton1,ResonantSoliton2} rather than a constant non-resonant pumping as considered here.

\begin{figure*}
  \includegraphics[width=0.85\textwidth]{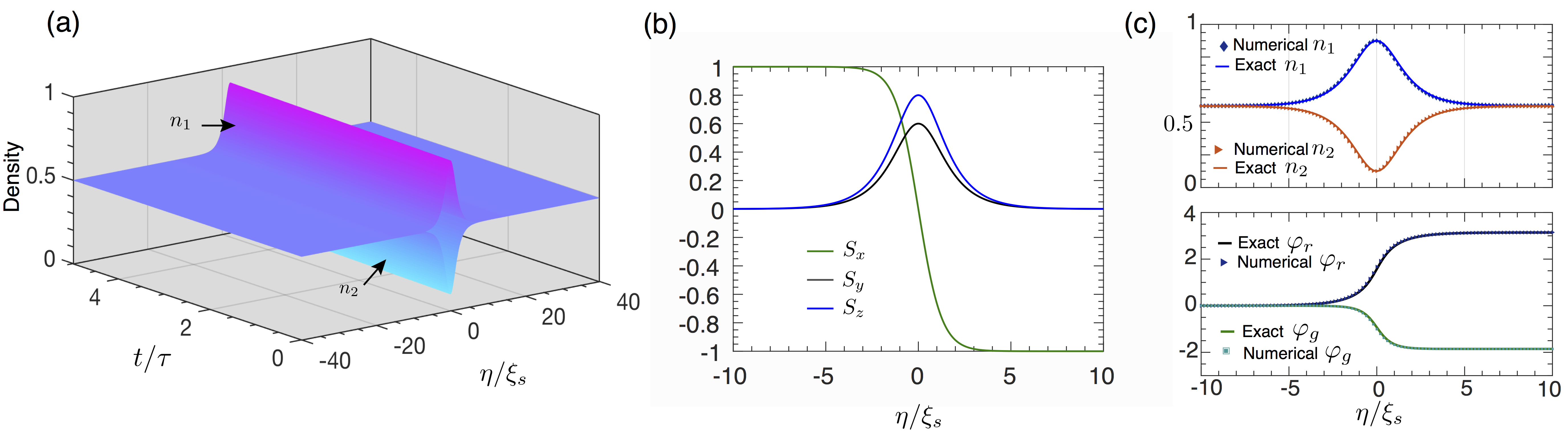}\\
\caption{Properties of a moving dissipative magnetic soliton. (a) Density distribution in space and time [see Eq.~(\ref{Spindensity})]. (b) Stokes parameters according to Eq. (\ref{eq:spin}). (c) Distribution of density of $n_1=|\psi_1|^2$ and $n_2=|\psi_2|^2$, relative phase $\varphi_{r}$ and global phase phase $\varphi_{g}$ at dimensionless time of $t/\tau=15$. In panel (c), exact analytical solutions are compared to numerical solutions from Eqs.~(\ref{GP1})-(\ref{Rate}) (see main text). In all plots, the parameters are chosen as $\gamma_C=0.01$ps$^{-1}$, $R=0.01$ps$^{-1}$$\mu m^2$, $g=0.01$meV$\mu m^2$, $P=0.41$ps$^{-1}$$\mu m^2$, $\gamma_R/\gamma_C=40$, $g_{12}/g=-0.1$, and $U=0.6$. 
}\label{figure2}
\end{figure*}

%as an \textit{exact} solution to the open-dissipative Gross-Pitaevskii (GP) 

%Although in optics we know few examples of the dissipative soliton limited in the density-channel as the exact solutions of the complex Ginzbrug-Landau equaiton~\cite{Akhmediev1995}, it's of interest to  explore the dissipative solitons in the spin-density channel in a controllable way. Moreover, our approach can be naturally extended to find the wanted types of vector dissipative solitons in the $N$-component dissipative GP equations.
We illustrate the central idea of our work by considering a spinor polariton BEC formed under uniform non-resonant pumping in a wire-shaped microcavity, as motivated by Ref.~\cite{Wertz2010}. The order parameter for the exciton-polariton BEC is a two-component complex vector~\cite{Borgh2010,Ohadi2015,Liew2015,Li2015,Askitopoulos2016}, which at quasi-1D is effectively described by ${\bf \psi}(x,t)\equiv [\psi_{1}(x,t),\psi_{2}(x,t)]^T$, where $\psi_{1}$ and $\psi_2$ are the spin-up and spin-down wavefunctions. The evolution of the order parameter is governed by the driven-dissipative GP equations coupled to the rate equation for the density $n_R$ of the reservoir polaritons~\cite{Pinsker2014,Pinsker2015,Pinsker2016,Xu2017,Xu2019}, i.e.,
\begin{eqnarray}
i\hbar\frac{\partial\psi_{1}}{\partial t}=\Big[-\frac{\hbar^{2}}{2m}\!\frac{\partial ^2}{\partial x^2}+g\left|\psi_{1}\right|^{2}&+&g_{12}\left|\psi_{2}\right|^{2}\Big]\psi_{1}\nonumber\\
&+&g_Rn_R\psi_{1}+{\it D}_s\psi_{1},\label{GP1}\\
i\hbar\frac{\partial\psi_{2}}{\partial t}=\Big[-\frac{\hbar^{2}}{2m}\!\frac{\partial ^2}{\partial x^2}\!+\!g\left|\psi_{2}\right|^{2}&+&g_{12}\left|\psi_{1}\right|^{2}\Big]\psi_{2}\nonumber\\
&+&g_Rn_R\psi_{2}+{\it D}_s\psi_{2},\label{GP2}\\
\frac{\partial n_{R}}{\partial t}=P-\Big[\gamma_{R}+R\Big(\left|\psi_{1}\right|^{2}&+&\left|\psi_{2}\right|^{2}\Big)\Big] n_{R}.\label{Rate}
\end{eqnarray}
Here, $m=10^{-4}m_e$ is the effective mass of polaritons ($m_e$ is the free electron mass), $g_{12}$ and $g$ are the interaction constants for opposite-spin and same-spin polaritons, respectively. The $V_R=g_Rn_R$ comes from the interaction between condensate and reservoir polaritons, with the interaction constant $g_R$ being spin-independent. The condensed polaritons with a finite lifetime $\gamma_C^{-1}$ are continuously replenished from reservoir polaritons at a rate $R$. This gain and loss process is captured by $D_s=i\hbar\left(Rn_{R}-\gamma_{C}\right)/2$.  The reservoir characterized by a decay rate $\gamma_R$ is driven by a spatially uniform and off-resonant continuous-wave (cw) pumping $P$. Hereafter we will denote by $n_{1(2)}=\left|\psi_{1(2)}\right|^{2}$ the density of spin-up (spin-down) components. Since for typical polaritonic systems $g>0$ and $g_{12}<0$ with $|g_{12}|\ll g$, the steady state condensate of above equations is linearly polarized with a stochastic polarization direction due to the absence of pinning~\cite{Rev0}. For collective excitations on top of the condensate, there are two excitation channels which correspond to the density $n=n_1+n_2$ and spin polarization $n_1-n_2$, respectively. We aim to show below that there exist exact vector soliton solutions $\psi_s\equiv[\psi_1^s,\psi_2^s]^T$ to Eqs.~(\ref{GP1})-(\ref{Rate}) under the condition
\begin{equation}
D_s\psi_{s}=0.\label{Darkstate}
\end{equation} 
As we will see, a key consequence of Eq.~(\ref{Darkstate}) is that the nonlinear dynamics of spin-polarization becomes decoupled and coherent, generating a localized soliton unaffected by the polariton loss, regardless of the coupling strength between the polariton BEC. 

%Furthermore, it's also of relevance to proposals and experiments for vector dissipative solitons in the context of the optics~\cite{VecDissipativeSoliton1,VecDissipativeSoliton2,VecDissipativeSoliton3,VecDissipativeSoliton4}.
To find the analytical expression of the traveling soliton solution $\psi_s(x-\upsilon t)$ with $
\upsilon $ being the velocity, we write the two-component order parameter $\psi$ in terms of densities and phases as following
\begin{equation}
\left(\begin{array}{c}
\psi_{1}\\
\psi_{2}
\end{array}\right)=\sqrt{\frac{n_0}{2}}\left(\begin{array}{c}
\sqrt{1+\delta n_1}e^{i\frac{\varphi_{r}}{2}}\\
\sqrt{1-\delta n_2}e^{-i\frac{\varphi_{r}}{2}}
\end{array}\right)e^{i\varphi_g/2}e^{-i\frac{\mu_Rt}{\hbar}}. \label{Msoliton}
\end{equation}
Here $n_0$ is assumed to be a constant density,  and $\mu_R=g_R\gamma_C/R$. The dimensionless variables $\delta n_{1,2}(x-\upsilon t)$ are defined from $n_{1}=n_0(1+\delta n_1)/2$ and $n_2=n_0(1-\delta n_1)/2$, respectively. In addition, $\varphi_{r}$ and $\varphi_g$ label the relative and global phases of $\psi_1$ and $\psi_2$, respectively, and we shall denote $\eta=x-\upsilon t$ for simplicity. Imposing boundary conditions $\lim_{\eta\rightarrow \pm\infty }\delta n_{1(2)}(\eta)=0$ and $\lim_{\eta\rightarrow \pm\infty }\partial_\eta\varphi_{r(g)}(\eta)=0$, we find~\cite{Supp} that Eqs.~(\ref{GP1})-(\ref{Darkstate}) can be exactly solved by Eq.~(\ref{Msoliton}) for $n_0=P/\gamma_{C}-\gamma_R/R$ and $n_R=\gamma_C/R$, and
\begin{eqnarray}
\delta n_1&=&\delta n_2=\sqrt{1-U^{2}}\text{sech}\Big[\frac{\eta}{\xi_s}\sqrt{1-U^{2}}\Big],\label{Spindensity}\\
\varphi_{r}&=&\arctan\Big[\frac{\sinh\left(\frac{\eta}{\xi_s}\sqrt{1-U^{2}}\right)}{U}\Big]+\frac{\pi}{2},\label{Rphase}\\
\varphi_{g}&=&-\text{arctan}\Big[\frac{\sqrt{1-U^{2}}\tanh\left(\frac{\eta}{\xi_s}\sqrt{1-U^{2}}\right)}{U}\Big]\nonumber\\
&-&\text{arctan}\Big[\frac{\sqrt{1-U^{2}}}{U}\Big].\label{Tphase}
\end{eqnarray}
Here $U\!\!=\!\!v/\sqrt{n_0(g-g_{12})/2m}$ is dimensionless velocity and $\xi_s\!\!=\!\!\hbar/\sqrt{2mn_0(g-g_{12})}$ is the spin healing length. 

In Fig.~\ref{figure2}(a), we illustrate the density distribution of a moving soliton with $U=0.6$, which preserves its shape during motion. The soliton profile at $t/\tau=15$ with $\tau=\hbar/(g-g_{12})n_0$ is shown in Figure~\ref{figure2}(c). Note $\lim_{\eta\rightarrow +\infty }\varphi_r-\lim_{\eta\rightarrow -\infty }\varphi_r=\pi$, i.e., the relative phase exhibits an exact $\pi$-jump, while for the global phase $\varphi_g$, the phase jump is not universal and depends on $U$ [see Fig.~\ref{figure2}(c)]. Further, we have numerically solved Eqs.~(\ref{GP1})-(\ref{Rate}), taking the initial order parameter given by Eqs.~(\ref{Msoliton})-(\ref{Tphase}) for $t=0$ along with $n_R(0)=\gamma_C/R$. Comparisons of numerical results at $t/\tau=15$ with analytical solutions show perfect agreement between the two [see Fig.~\ref{figure2} (c)]. We have further numerically verified the stability of the soliton by time-evolving an initial order parameter where $n_1(0)-n_2(0)$ is perturbed from Eq.~(\ref{Spindensity}) while keeping $n_0=n_1(0)+n_2(0)$ fixed. 

The vector soliton solution given by Eqs.~(\ref{Msoliton})-(\ref{Tphase}) represents a localized spin polarization residing on top of a linearly polarized background state (i.e., a magnetic soliton). In characterizing the polarization property of the soliton, we use the experimentally accessible Stokes parameters~\cite{Shelykh2006,Ohadi2015,Sich2018}. The distribution of the linear and circular polarization degrees of the soliton are defined by 
\begin{equation}
S_x(\eta)=\frac{2\Re{(\psi^*_1\psi_2)}}{n_0}, \hspace{2mm} S_z(\eta)=\frac{(|\psi_1|^2-|\psi_2|^2)}{n_0}. \label{eq:spin}
\end{equation}
As illustrated in Fig.~\ref{figure2}(b)~\cite{Supp}, the moving soliton is strongly elliptically polarized for $\eta<l_w$ with $l_w=\xi_s/\sqrt{1-U^2}$ being the soliton width. Right at the center, $S_z(\eta)=\sqrt{1-U^2}$ is maximum. Far from the center, the soliton becomes linearly polarized with equal occupation in spin-up and spin-down components, i.e., $\lim_{\eta\rightarrow \pm \infty} n_1=n_2=n_0/2$. But the linear polarization degree flips its direction across the soliton profile from one side to the other due to the $\pi$ jump in $\varphi_r$. The total spin polarization degrees of the soliton is $\int d\eta S_z(\eta)=\pi\xi_s$ independent of the velocity $U$. Since this magnetic soliton features spin-balanced density background, it differs from other vector solitons in polaritonic systems, such as dark-bright solitons~\cite{Xu2019}, where the background state is spin-polarized. 

%This is the key difference between the vector dissipative soliton and the soltion solutions of the integrable Hamiltonian systems~\cite{Qu2016}, which usually have the amplitude as a free parameter. 

Remarkably, when moving in the dissipative system, this magnetic polariton soliton preserves its energy, i.e., it is a dissipative soliton. Following Refs.~\cite{Kivshar1994,Smirnov2014,Xu2019}, the energy $E$ of the soliton is calculated according to
\begin{eqnarray}
E&=&\int dx{\psi}^\dag\left(-\frac{\hbar^2}{2m}\frac{\partial ^2}{\partial x^2}\right)\psi+\frac{g-g_{12}}{4}\int dx[n_1-n_2]^2\nonumber\\
&+&\frac{g+g_{12}}{4}\int dx\left[n_1+n_2-n_0\right]^2.\label{Ham}
\end{eqnarray}
Using Eqs.~(\ref{GP1})-(\ref{Rate}), along with soliton solutions (\ref{Msoliton})-(\ref{Tphase}), we obtain~\cite{Supp}
\begin{eqnarray}
\frac{d E}{d t}=-2\Re{\int\left[D_s\left(\frac{\partial {\psi^*_1}}{\partial t}+\frac{\partial {\psi^*_2}}{\partial t}\right)\right]dx}=0, \label{Energy}
\end{eqnarray}
where $\Re$ denotes the real component, and $D_s$ is defined in Eqs.~(\ref{GP1})-(\ref{GP2}). 

Dissipative magnetic soliton in Eqs.~(\ref{Msoliton})-(\ref{Tphase}) features a constant $n_0$ across the soliton profile. This is enabled in our case by the stable balance between the gain and loss which maintains the density of the condensed polariton at $n_0=P/\gamma_{C}-\gamma_R/R$ for given system parameters. Thanks to Eq.~(\ref{Darkstate}), on the other hand, one can obtain a \textit{closed real} equation for circular polarization degrees, i.e., $\left[d S_z(y)/dy\right]^{2}+S_z^{4}(y)-\left(1-U^{2}\right)S_z^{2}(y)=0$ with $y=\eta/\xi_s$. The solution to this equation [i.e., Eq.~(\ref{Spindensity})] therefore results directly from the competition between the kinetic energy and the nonlinear interaction. The double balance picture undelying dissipative magnetic soliton is in crucial contrast to the well known dissipative Kerr soliton solution~\cite{DissipativeSolitonRev1} of the complex Ginzburg-Landau equation. There, the composite balances take place in a single excitation channel, whereas dissipative magnetic soliton stems from separate balance in two decoupled channels, i.e., the balance of gain and loss fixing background density, and the balance of nonlinearity and dispersion in the spin-polarization channel dictates soliton profile. In polariton condensates, other dissipative solitons have been reported~\cite{ResonantSoliton1,ResonantSoliton2,Xue2014,Pinsker2014,Ma2017,Ma2018}, which form from single-channel double balance (mostly in density channel), and either require spatially dependent pumping or resonant pumping, while a uniform nonresonant pumping is used here.

\begin{figure}
  \includegraphics[width=1.0\columnwidth]{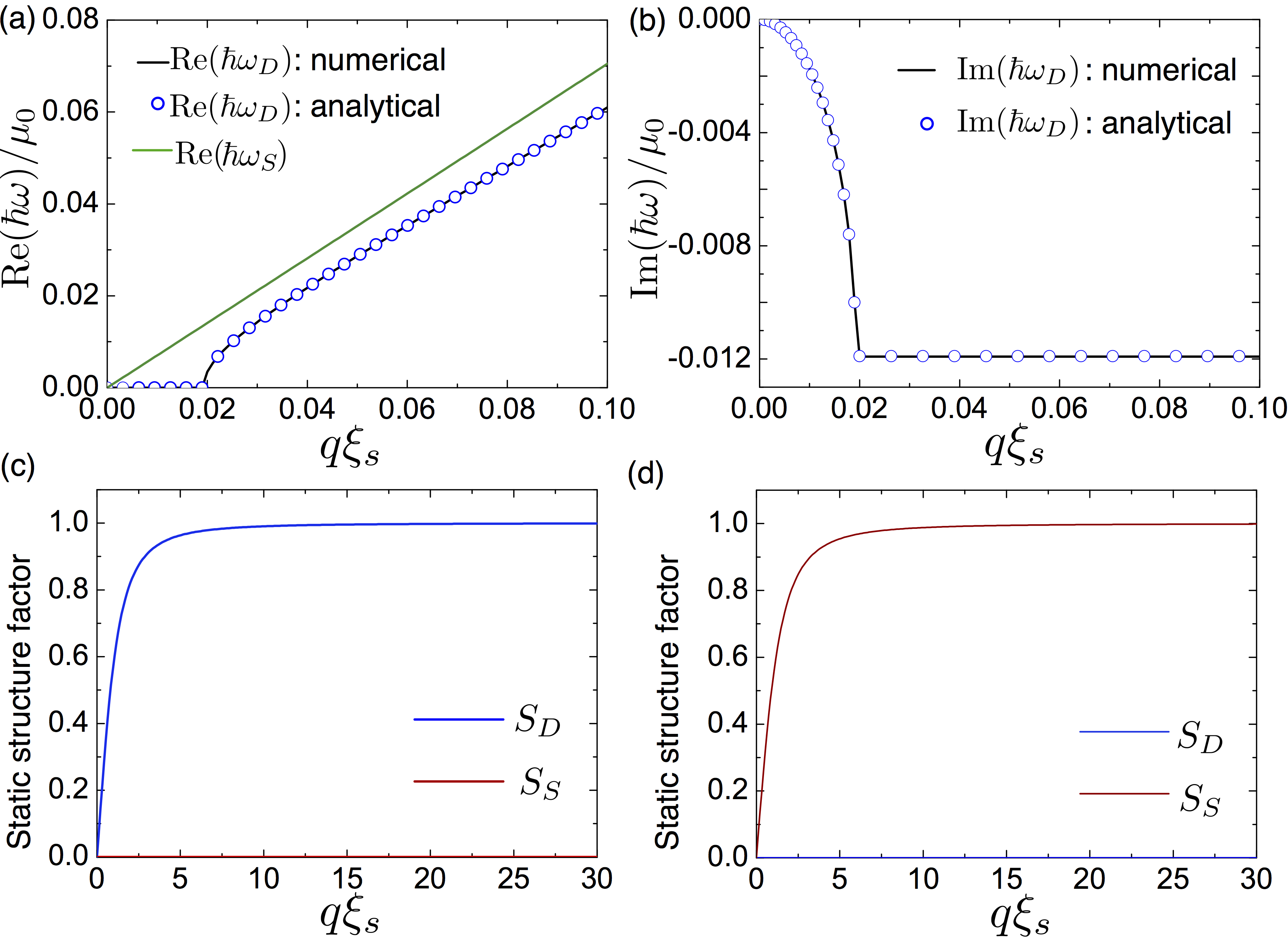}\\
\caption{Visualization of the decoupling of spin-polarization channel from the density channel through linear excitations. Panels (a) and (b): (a) real and (b) imaginary parts of the energy of density excitation $\hbar\omega_D$ and energy $\hbar\omega_S$ of spin-polarization excitation. Solid curves depict numerical solutions of Bogoliubov's equations, and the curve with circles indicate analytical solutions. Panels (c)-(d): Density static structure factor $S_D(q)$ and spin-density static structure factor $S_S(q)$ when the spinor polariton BEC is subjected to a perturbation in the form (c) $\lambda e^{i(qx-\omega t)}+\textrm{H.c}$ and (d) $\lambda \sigma_z e^{i(qx-\omega t)}+\textrm{H.c}$ (see main text). In all plots, we have used the same parameters as in Fig.~\ref{figure2}. 
}\label{figure3}
\end{figure}
We note that exact solutions of the form in Eqs.~(\ref{Spindensity})-(\ref{Tphase}) are previously known to exist in equilibrium atomic two-component BECs~\cite{Qu2016,Qu2017,Danaila2016,Congy2016} under the condition $g-g_{12}\ll g$. There, the soliton is purely sustained by the nonlinear effect compensating dispersion, and relies on the ''energetic mechanism'' to physically ensure an (approximately) unperturbed $n_0$: When $g-g_{12}\ll g$, creating considerable density depletion will cost substantial energy, and is thus strongly suppressed even at the center of the soliton. Instead, dissipative magnetic soliton is inherently non-equilibrium and relies on a ''gain and loss'' mechanism to fix $n_0$, thus it can exist beyond the condition of $g-g_{12}\ll g$, such as here in polaritonic systems where typically $g >0$ and $g_{12}<0$~\cite{Rev0,Rev1,Rev2,Rev3}.

To gain more physical understanding that nonlinear spin-polarization dynamics is decoupled from other excitation channels and is coherent, which is crucial for the non-decaying property of the magnetic polariton soliton, we note that linear excitations can be viewed as building blocks of nonlinear excitation. In this light, it is beneficial to analyze the properties of linear spin polarization excitation, such as the spectrum and linear response function which are observable, as we describe below and detail in supplementary material~\cite{Supp}. To describe a spinor polariton BEC perturbed from the steady state in the linear regime, we can substitute Eq.~(\ref{Msoliton}) into Eqs.~(\ref{GP1})-(\ref{Rate}) and follow the standard Bogoliubov-de Gennes (BdG) approach. The resulting equation for eigenenergy $\hbar\omega_{{q}}$ of excitations is $[\left(\hbar\omega_{{q}}\right)^{2}-\left(\hbar\omega_{S}\right)^{2}]\times\{ \left(\hbar\omega_{{q}}\right)^{3}+i\left(Rn_{0}+\gamma_{R}\right)\left(\hbar\omega_{{ q}}\right)^{2}-[Rn_{0}\gamma_{C}+\left(\hbar\omega_{B}\right)^{2}]\hbar\omega_{{q}}+ic\left({q}\right)\} =0$, where $\hbar\omega_S=\sqrt{\varepsilon_{{q}}^{0}\left[\varepsilon_{q}^{0}+\left(g-g_{12}\right)n_{0}\right]}$, $\hbar\omega_{B}=\sqrt{\varepsilon_{q}^{0}\left[\varepsilon_{q}^{0}+\left(g+g_{12}\right)n_{0}\right]}$, and $c(q)=-(Rn_0+\gamma_R)(\hbar\omega_B)^2+2gn_0\gamma_c\varepsilon_q^0$,  with $\epsilon_q^0=\hbar^2q^2/(2m)$ being the free-particle energy. This obviously entails two decoupled equations: the quadratic equation immediately yields $\hbar\omega_{{q}}=\pm \hbar\omega_S$ for the energy of the spin-polarization excitation, whereas the cubic equation reflects the coupled linear excitations in the reservoir and density channel of polariton BEC.  Importantly, we see that $\omega_s$ is purely \textit{real}, regardless the reservoir is fast or slow compared to the polariton BEC; see Fig.~\ref{figure3}(a). This feature of the linear spin polarization excitation is in contrast to the linear density excitation which generically exhibits a complex energy $\hbar\omega_{D}$ and eventually damps out, as can be transparently seen in the fast reservoir limit $\gamma_R/\gamma_C\gg 1$. There, adiabatic elimination of the reservoir enables a simple expression $\hbar\omega_D=-i\Gamma/2\pm \hbar \omega_0$, with $\hbar\omega_0=\sqrt{\varepsilon_{q}^{0}\left[\varepsilon_{q}^{0}+\left(g+g_{12}\right)n_{0}-2g_{R}\hbar\Gamma/R\right]-\Gamma^{2}/4}$ and $\Gamma=n_{0}n_{R}^{0}R^{2}\hbar/(\text{\ensuremath{\gamma_{R}}}+n_{0}R)$. Note $\omega_D$ is purely imaginary for $|q|\le q_c$ due to the polariton loss, with $q_{c}=\sqrt{m(\sqrt{\alpha^{2}+\Gamma^{4}/4}-\alpha^2)/\hbar^{2}}$. In Figs.~\ref{figure3}(a) and (b), we have plotted the complex spectrum of linear density excitation in the fast reservoir limit, comparing analytical results with numerical solutions of the BdG equations. The two agree well with each other. 

Turning next to the linear response, suppose the spin polariton BEC is subjected to an external density perturbation standardly described by $\lambda e^{i(qx-\omega t)}+\textrm{H.c}$ with $\lambda\ll 1$. We are interested in the system response characterized by the density static structure factor $S_{D}({q})$ and the spin-density static structure factor $S_S(q)$~\cite{PineBook}. For simplicity, we assume fast reservoir limit and analytically derive $S_{D}({q})$ as~\cite{Supp}
\begin{eqnarray}
S_{D}({ q})=\begin{cases}
\begin{array}{c}
\frac{\varepsilon_{{\bf q}}^{0}}{\pi\hbar\left|\hbar\omega_{0}\right|}\log\left(\frac{\Gamma+2\left|\hbar\omega_{0}\right|}{\Gamma-2\left|\hbar\omega_{0}\right|}\right)\\
\frac{4\varepsilon_{{\bf q}}^{0}}{\pi\hbar\Gamma}\\
\left[\frac{1}{2}+\frac{1}{\pi}\tan^{-1}\left(\frac{4\omega_{0}^{2}-\Gamma^{2}}{4\Gamma\ensuremath{\omega_{0}}}\right)\right]\frac{\epsilon_{{\bf q}}^{0}}{\hbar\omega_{0}}
\end{array} & \begin{array}{c}
{q<{ q_{c}}},\\
{q}={ q_{c}},\\
{ q}>{ q_{c}}
\end{array}\end{cases}\label{DSF}
\end{eqnarray}
We also find $S_S(q)=0$. As shown in Fig.~\ref{figure3}(c), we have $\lim_{{q}\rightarrow \infty} S_D(q)\rightarrow 1$, with physical implication that the linear response of polariton BEC under a density perturbation is exhausted by density excitation, without generating excitations in the spin polarization channel. If the system is instead subjected to a spin-dependent perturbation $\lambda \sigma_z e^{i(qx-\omega t)}+\textrm{H.c}$, we can derive $S_{S}(q)=\hbar{q}^{2}/(2m\omega_S)$ which approaches unity for ${q}\rightarrow \infty$ and $S_D(q)=0$ [see Fig.~\ref{figure3}(d)]. This further corroborates that a spin polarization perturbation only induces spin excitations. With above spectrum and linear response analysis, we conclude the linear spin-polarization excitation on top of the linearly-polarized condensate in our system comprises a closed channel unaffected by the loss effect, providing the base for forming a non-decaying nonlinear excitation.

In summary, we have introduced dissipative magnetic polariton soliton representing an exact solution to driven-dissipative two-component GP equations, which manifests a bi-channel double balance that can serve as a new scenario for forming dissipative solitons in generic multi-component dissipative nonlinear systems. This result further enriches our understanding of the vector matter-light solitons, and can be of interest along the line of ultrafast information processing, where polariton solitons have been identified as promising candidates due to their picosecond response time. While in our illustration the condition of Eq.~(\ref{Darkstate}) reduces to $D_s=0$, it will be interesting in the future to extend the underlying essential idea to achieve non-decaying solitons in more generic cases where $D_s\psi_s=0$ rather than $D_s=0$ holds. In a broader context, multi-component dissipative nonlinear system are widely seen, including mode-locked lasers and optical microresonators~\cite{DissipativeSoliton1,DissipativeSoliton2}. The bi-channel double mechanism reported in this work and its variants may lead to new dissipative solitons in these systems.

{\it Acknowledgement---} We acknowledge constructive suggestions from Augusto Smerzi, and thank Xingran Xu, Biao Wu, Chao Gao, and Yan Xue for stimulating
discussions. This work is financially supported by the key projects of the Natural Science Foundation of China (Grant No. 11835011) and the Natural Science Foundation of China (Nos. 11874038, 11434015, 61835013). Y.H. also acknowledges support from the National Thousand-Young-Talents Program, and Changjiang Scholars and Innovative Research Team (Grant No. IRT13076). W.M.L. is also supported by the National Key 
R$\&$D Program of China (Grant No. 2016YFA0301500) and the Strategic Priority Research Program of the Chinese Academy of Sciences (Grant Nos. XDB01020300, XDB21030300).
\bibliography{myr}

\end{document}